\documentclass[aps,prl,showpacs,superscriptaddress,twocolumn]{revtex4}

\usepackage{mathrsfs}
\usepackage{amsfonts}
\usepackage{amssymb}
\usepackage{amsmath}
\usepackage{graphicx}
\usepackage{sidecap}
\usepackage{color}
\usepackage[colorlinks=true,citecolor=blue,linkcolor=magenta]{hyperref}
\usepackage{subeqnarray}
\usepackage{verbatim}
\usepackage{ulem}

\newcommand{\ket}[1]{\vert #1 \rangle}

\begin{document}

\title{Measurement-Based Quantum Computation on Two-Body Interacting Qubits with Adiabatic Evolution}

\author{Thi Ha Kyaw}
\affiliation{Centre for Quantum Technologies, National University of Singapore, 3 Science Drive 2, Singapore 117543, Singapore}

\author{Ying Li}
\affiliation{Centre for Quantum Technologies, National University of Singapore, 3 Science Drive 2, Singapore 117543, Singapore}
\affiliation{Department of Materials, University of Oxford, Parks Road, Oxford OX1 3PH, United Kingdom}

\author{Leong Chuan Kwek}
\affiliation{Centre for Quantum Technologies, National University of Singapore, 3 Science Drive 2, Singapore 117543, Singapore}
\affiliation{Institute of Advanced Studies, Nanyang Technological University, 60 Nanyang View, Singapore 639673, Singapore}
\affiliation{National Institute of Education, Nanyang Technological University, 1 Nanyang Walk, Singapore 637616, Singapore}

\date{\today}

\begin{abstract}
A cluster state cannot be a unique ground state of a two-body interacting Hamiltonian.
Here, we propose the creation of a cluster state of logical qubits encoded in spin-1/2 particles by adiabatically weakening two-body interactions.
The proposal is valid for any spatial dimensional cluster states.
Errors induced by thermal fluctuations and adiabatic evolution within finite time can be eliminated ensuring fault-tolerant quantum computing schemes.
\end{abstract}

\maketitle

\textit{Introduction}.--- Measurement-based quantum computation (MBQC) \cite{MBQC} provides a promising alternative paradigm of quantum information processing where computation is done through single-particle measurements on some highly entangled resource states \cite{Nest2006}.
As long as these highly entangled resource states are available, no entangling operation is ever needed.
Ideally, it is desirable to prepare these highly entangled resources through appropriate engineering of thermal ground states in natural physical systems.
One resource for the MBQC is the cluster state \cite{Briegel2001} which is the ground state of spin-1/2 particles with $k$-body interactions where $k\ge 3$ \cite{5body}.
Unfortunately, cluster states are not the exact unique ground state of any Hamiltonian with only two-body interactions \cite{Nielsen2005}.

Cluster states are obtained from ground states of two-body Hamiltonians with spin$> 1/2$ e.g., the one-dimensional AKLT model of spin-1 particles \cite{Brennen2008} or with higher dimension, e.g., the two-dimensional AKLT model of spin-3/2 particles \cite{2DAKLT} under appropriate projective measurements.
It should be noted that as soon as the system cools to the ground state, any interaction between the particles should be switched off immediately to avoid further degradation of quantum correlations needed for the quantum processing \cite{Brennen2008,2DAKLT,2body}, except for the case where the system evolves under always-on periodically driven interactions \cite{Li2011}.
Sometimes, such an interaction may need to be switched off adiabatically in order to isolate the quantum information encoded in the edge states \cite{Miyake2010}.

In this Letter, we look at MBQC on two-body interacting spin-1/2 particles subject to adiabatic evolution. Two-body spin-1/2 interactions are generally preferred as such systems, compared to systems with higher spins and many-body interactions, are generally better suited for experimental implementation \cite{Kim2010}. Our proposal is closer in spirit to a previous work of Ref. \citenum{FTQC1}, where cluster states are created with only nearest-neighbor Ising-type interactions, which have been experimentally realized with neutral atoms in optical lattices \cite{cluster_exp}. On the other hand, a five-body interaction for the cluster state can also be obtained effectively from two-body interactions via perturbations \cite{Stephen_2006}. To obtain an approximate cluster state, this perturbation should be sufficiently weak. However, a weak perturbation implies a small energy gap between the ground state and the first excited state. This naturally leads to the need for the system to be cooled to a sufficiently low temperature depending on the size of the energy gap. Here, we create cluster states by adiabatically evolving the ground state of two-body interacting spin-1/2 particles with a built-in energy gap protection. Thus, we hope to operate our proposed systems at a higher temperature environment.

The principal motivation behind our proposal stems from adiabatic quantum computing (AQC) \cite{Farhi_2000}.
However, there is a key difference between AQC and our proposal.
In AQC, a system is initially prepared in the ground state of a simple Hamiltonian.
By adiabatically interpolating the Hamiltonian to a target Hamiltonian, a desired state is obtained as the final state in accordance with the adiabatic theorem.
In addition, the ground states of instantaneous Hamiltonians usually need to be protected by a finite energy gap throughout the entire evolution \cite{Sarandy2005}.
To create a cluster state with the standard AQC scheme, we need a target Hamiltonian whose unique ground state is the cluster state.
While a cluster state is never a unique ground state of any two-body interaction Hamiltonian, it can still be one of the degenerate ground states.
By slowly weakening the interactions of a two-body Hamiltonian, we show that the system could eventually achieve a cluster state as one of the degenerate ground states.
It should also be emphasized that even though our proposal can yield a cluster state of logical qubits as the target state, the energy gap disappears at the end of the adiabatic evolution.
Fortunately, thanks to the inherent symmetry of stabilizers, the desired ground state is protected from the noise due to the finite speed of evolution even if the energy gap vanishes.
In this sense, our proposal differs from the standard AQC schemes.

In our models, we regard each qubit of the cluster state as a logical qubit of several spin-1/2 particles.
The quantum correlations, i.e., \textit{stabilizers}, of the cluster state are established in the initial state, which is the ground state protected by a large energy gap and therefore tolerates a relatively higher temperature.
The initial state cannot be used for the MBQC, since it is outside the logical subspace where logical qubits are encoded.
These cluster-state correlations are preserved during the adiabatic evolution.
In this way, the final state is a cluster state of a logical qubit, which can be converted into a cluster state of spin-1/2 particles via single-qubit measurements.

\textit{The general protocol}.---We encode each qubit of the cluster state in $n$ spin-1/2 particles as
\begin{equation}
\ket{0}_j = \otimes _{m=1}^n \ket{\uparrow}_{j,m},
\hspace{0.5cm}
\ket{1}_j = \otimes _{m=1}^n \ket{\downarrow}_{j,m}.
\end{equation}
Here, the $j$th logical qubit is encoded in spin-1/2 particles $\{ (j,m) : m=1,2,\ldots,n \}$, and $\ket{\uparrow}_{j,m}$ ($\ket{\downarrow}_{j,m}$) is the eigenstate of the Pauli operator $\sigma ^z_{j,m}$ with the eigenvalue $+1$ ($-1$).
These logical states are stabilized by operators $\{ \sigma ^z_{j,1}\sigma ^z_{j,m}\}$, i.e., logical states are common eigenstates of these operators with eigenvalue $+1$.
Pauli operators of the $j$th logical qubit are
\begin{equation}
X_j = \prod _{m=1}^n \sigma ^x_{j,m}
\hspace{0.2cm} \text{and} \hspace{0.2cm}
Z_j = \sigma ^z_{j,1}.
\end{equation}
This encoding has been used for constructing a perturbative model of the cluster state \cite{Stephen_2006}.

The cluster state is the common eigenstate with eigenvalue $+1$ of cluster-state stabilizers \cite{MBQC}
$
S_j = X_j\prod _{i\in nb(j)}Z_i = \prod _{m=1}^n \sigma ^x_{j,m} \prod _{i\in nb(j)} \sigma ^z_{i,1},
$
where $nb(j)$ is the set of nearest neighboring logical qubits of the $j$th logical qubit.
Hence, on the spin-1/2-particle level, the cluster state is stabilized by $\{S_j\} \cup \{\sigma ^z_{j,1}\sigma ^z_{j,m}\}$.
By noticing that a product of stabilizers is also a stabilizer, cluster-state stabilizers can be rewritten as
$
S_j^{\{ m_{j,i} \}} = S_j\prod _{i\in nb(j)}\sigma ^z_{i,1}\sigma ^z_{i,m_{j,i}} 
= \prod _{m=1}^n \sigma ^x_{j,m} \prod _{i\in nb(j)} \sigma ^z_{i,m_{j,i}},
$
where $\{ m_{j,i} \}$ is a string of numbers satisfying $1\leq m_{j,i}\leq n$.
If a state is stabilized by $\{S_j^{\{ m_{j,i} \}}\} \cup \{\sigma ^z_{j,1}\sigma ^z_{j,m}\}$ for any choice of $\{ m_{j,i} \}$, the state is the cluster state.
This cluster state of logical qubits can be converted into a cluster state of physical qubits by measuring $\sigma^x$ of arbitrary $n-1$ physical qubits of each logical qubit.
Therefore, this cluster state of logical qubits is a universal resource for the MBQC.

To obtain the cluster state via adiabatic cluster-state concentration, we consider a Hamiltonian of $N\times n$ spin-1/2 particles in the form
\begin{equation}
H=H_0+\lambda V,
\label{HG}
\end{equation}
where $H_0 = \sum _{j=1}^N h_j$ is a Hamiltonian of Ising interactions,
$
h_j = -J\sum _{m=1}^n \sigma ^z_{j,m} \sigma ^z_{j,m+1},
$
where $\sigma ^z_{j,n+1} = \sigma ^z_{j,1}$, and $J$ is the coupling constant of Ising interactions.
Here, $V$ denotes some two-body interactions that satisfies the following conditions:
i) $V$ commutes with a set of cluster-state stabilizers $\{S_j^{\{ m_{j,i} \}}\}$ corresponding to one choice of $\{ m_{j,i} \}$;
and ii) when the interaction strength $\lambda$ is nonzero, degenerate ground states are split. As the result, $H$ has a unique ground state with a finite energy gap above it.
Our protocol of cluster-state concentration includes two steps: 1) cooling the system with a nonzero $\lambda$ to the ground state; 2) adiabatically switching off $\lambda$.
In the adiabatic limit, the final state is the cluster state of logical qubits up to some single-particle Pauli operations.

This protocol relies on the set of cluster-state stabilizers $\{S_j^{\{ m_{j,i} \}}\}$ that are conserved quantities for any value of $\lambda$, i.e., $[ H, S_j^{\{ m_{j,i} \}} ]=0$, $\forall \lambda$.
We would like to remark that $H_0$ commutes with $S_j^{\{ m_{j,i} \}}$.
Hence, the unique ground state of $H$ for any nonzero $\lambda$ is the common eigenstate of cluster-state stabilizers.
We suppose corresponding eigenvalues are $\{s_j^{\{ m_{j,i} \}}\}$, where $s_j^{\{ m_{j,i} \}} = +1$ or $-1$.
Therefore, if the initial state is the ground state with a nonzero $\lambda$, the final state is still a common eigenstate of cluster-state stabilizers with the same eigenvalues.

For each logical qubit, $\ket{0}_j$ and $\ket{1}_j$ are degenerate ground states of $h_j$.
The ground-state subspace of $H_0$ is $2^N$-fold degenerate, which coincides with the subspace encoding logical qubits.
During the adiabatic evolution, the state always remains in the ground state of the instantaneous Hamiltonian  \cite{Farhi_2000}. Thus, the final state is in the ground-state subspace of $H_0$, i.e. in the logical subspace.
Any state in the logical subspace is stabilized by $\{ \sigma ^z_{j,1}\sigma ^z_{j,m}\}$.
Therefore, the final state is the common eigenstate of $\{S_j^{\{ m_{j,i} \}}\}$ and $\{\sigma ^z_{j,1}\sigma ^z_{j,m}\}$ with eigenvalues $\{s_j^{\{ m_{j,i} \}}\}$ and $\{+1\}$, respectively.
By performing single-particle Pauli operations $[(1+s_j^{\{ m_{j,i} \}})\openone+(1-s_j^{\{ m_{j,i} \}})\sigma ^z_{j,1}]/2$, the final state can be transformed into the cluster state of logical qubits.

When $\lambda$ adiabatically approaches zero , the energy gap between the ground state and first-excited state vanishes, which usually implies one has to slow down the rate of change of $\lambda$ to avoid any inadvertent excitation.
Fortunately, in the degenerate subspace, i.e., the logical subspace, the cluster state is the only state with eigenvalues $\{s_j^{\{ m_{j,i} \}}\}$.
Similarly, the ground state is the only state with eigenvalues $\{s_j^{\{ m_{j,i} \}}\}$ in all states split from the degenerate subspace.
Therefore, the transitions between the ground states and other states split from the degenerate subspace are forbidden; i.e., one does not have to slow down the rate of change of $\lambda$, according to the vanishing energy gap, when $\lambda \rightarrow 0$.

\begin{figure}[t]
\centering
\includegraphics[scale=0.6]{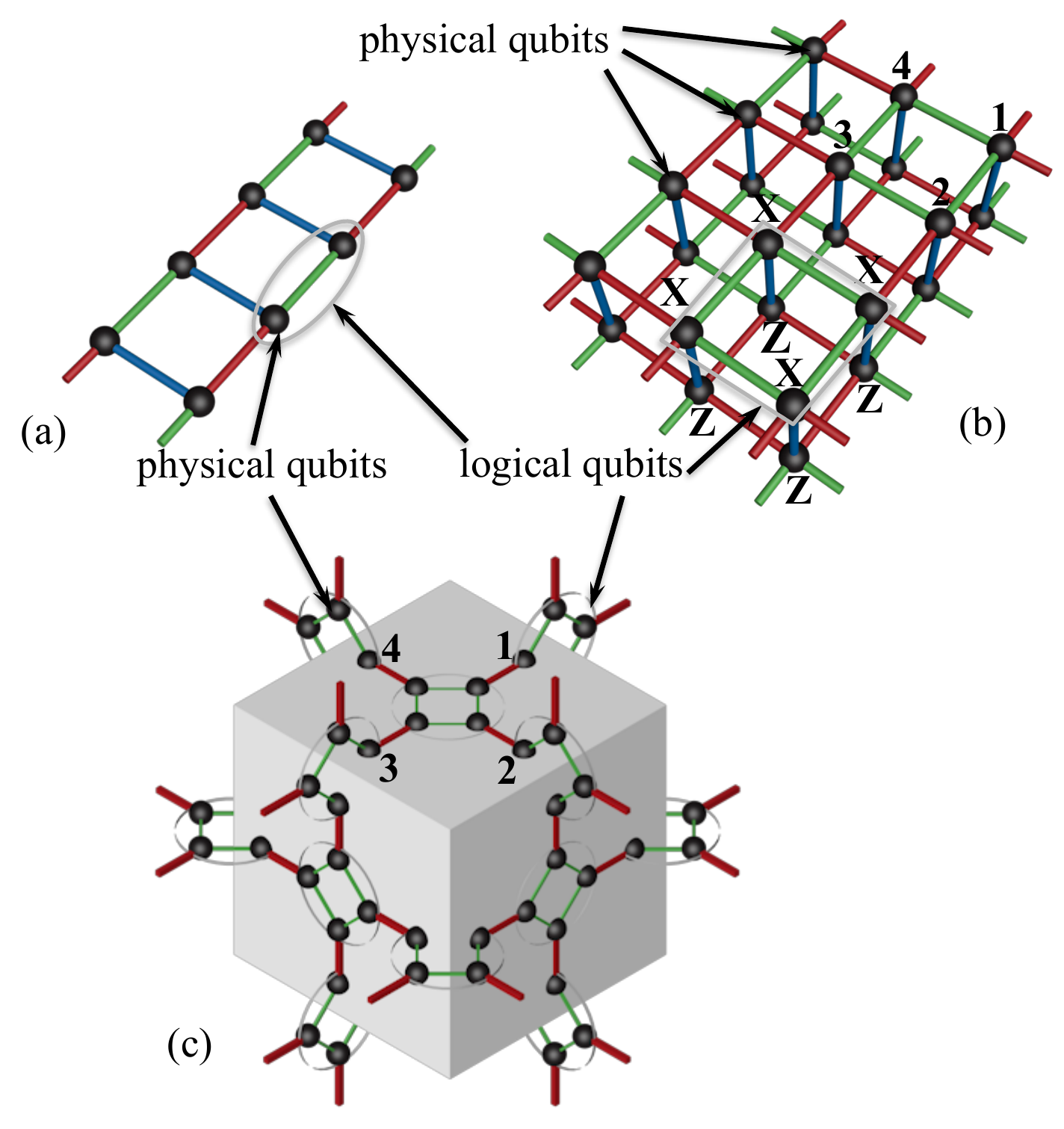}
\caption{
In all the figures, black circles represent spin-1/2 particles, red bonds denote $\sigma^x_i \sigma^x_j$ interactions, blue bonds denote $\sigma^y_i \sigma^y_j$ interactions, and green bonds are $\sigma^z_i \sigma^z_j$ interactions where $i$ and $j$ are labels of two corresponding spin-1/2 particles.
(a) One-dimensional Kitaev model.
A grey ellipse of two spin-1/2 particles connected by a green bond represents a logical qubit.
(b) Two-dimensional Kitaev-like model.
A grey square of four spin-1/2 particles connected by four green bonds represents a logical qubit.
Numbers 1, 2, 3 and 4 label the four physical qubits that belong to the logical qubit $j$ (the grey square).
(c) An elementary cubic lattice in a three-dimensional square lattice model.
A grey circle of four spin-1/2 particles connected by four green bonds denotes a logical qubit.
Numbers 1, 2, 3 and 4 label the four logical qubits located at four edges surrounding the central $j$th logical qubit at the top face of the cube.
}
\label{models}
\end{figure}

\textit{1D Kitaev model}.---We illustrate our protocol using the one-dimensional Kitaev model, shown in Fig. \ref{models}(a), and how it can be used to create a one-dimensional cluster state.
The one-dimensional cluster state is not a resource state for universal MBQC, but it is still useful for implementing single qubit gates \cite{MBQC} or for transferring a quantum state through a quantum wire \cite{Gross_2007}.

The Hamiltonian of the one-dimensional Kitaev model is
$
H^{\rm 1D} = H_0 ^{\rm 1D} +\lambda V^{\rm 1D}, 
$ where
$
H_0 ^{\rm 1D} = -J\sum_j \sigma^z _{j,1} \sigma^z _{j,2}, 
$ and
$V^{\rm 1D} = - \sum_j (\sigma^x_{j,1} \sigma^x_{j-2,2} + \sigma^y_{j,1} \sigma^y_{j-1,2}).
$
Here, each logical qubit is encoded in a pair of spin-1/2 particles.
The ground state is nondegenerate with a finite energy gap when $0<\lambda<J/2$ \cite{Kitaev_2006}.
We note that, for each plaquette $j$ [see Fig. \ref{models}(a)], there is a conserved quantity
\begin{equation}
W_j =\sigma_{j,1}^x \sigma_{j,2}^x \sigma_{j-1,2}^z \sigma_{j+1,1}^z,\label{1D_stabilizer}
\end{equation}
i.e., $[H^{\text{1D}},W_j]=0$.
The conserved quantity is given by $W_j=S_j^{\{ m_{j,i} \}}$ with $m_{j,j-1}=2$ and $m_{j,j+1}=1$.
Therefore, the Hamiltonian $H^{\rm 1D}$ satisfies the form Eq. (\ref{HG}).
Since $W_j$'s commute with each other, they can be diagonalized simultaneously with eigenvalues $w_j =\pm 1$, thus allowing us to partition the total Hilbert space into invariant subspaces of $H^{\rm 1D}$.
It has been shown \cite{vortex_free} that the ground state lies in the vortex-free subspace with $w_j =+1$, $\forall j$.
Since $W_j$'s are conserved quantities, the evolution is restricted to the vortex-free subspace.
Therefore, the energy gap that protects the adiabatic evolution is always nonzero.
The excitation spectrum is obtained by mapping \cite{Chen2008} the original Hamiltonian $H^{\text{1D}}$ into $p$-wave Fermi superfluid representation where the energy gap between the ground state and the first excited state is $\Delta E^{\text{1D}} = 2J-4\lambda$.

With this energy gap, we first cool our system to its ground state with nonzero $\lambda$ as elaborated in the general protocol section.
We then adiabatically switch off $\lambda$ so that the final state is in the ground-state subspace of $H_0 ^{\text{1D}}$, stabilized by $W_j =X_j Z_{j-1} Z_{j+1}$, Eq. \ref{1D_stabilizer}, yielding our 1D cluster state.

{\it 2D Kitaev-like model}.---To qualify for a resource state of the universal MBQC, it has to be at least two dimensional.
Motivated by the one-dimensional cluster state, we propose a two-dimensional Kitaev-like model [see Fig. \ref{models}(b)] that satisfies the form of Eq. (\ref{HG}).
We show here that a two-dimensional Kitaev-like model can be used to create a two-dimensional cluster state.

The Hamiltonian of the two-dimensional model reads $H^{\rm{2D}}= H_0 ^{\rm 2D} + \lambda V^{\rm 2D}$, where
$
H_0 ^{\rm 2D}= -J \sum_{\bf j} \sum_{\langle \mu,\mu' \rangle}
\sigma^z _{{\bf j},\mu} \sigma^z _{{\bf j},\mu'}
$
and
$
V^{\rm 2D} = - \sum_{\langle {\bf j},{\bf j}' \rangle} \sum_\mu
( \sigma^x _{{\bf j},\mu} \sigma^x _{{\bf j}',\mu_r}
+ \sigma^y _{{\bf j},\mu} \sigma^y _{{\bf j}',\mu_b} ).
$
Each logical qubit here is encoded in four spin-1/2 particles: ${\bf j} =(j_1,j_2)$ is the coordinate of a logical qubit, $\langle \mu,\mu' \rangle$ labels two connected spin-1/2 particles which belong to the same logical qubit located at $\bf j$, $\langle {\bf j},{\bf j}' \rangle$ denotes two connected logical qubits, and $\mu_r$ ($\mu_b$) denotes the spin-1/2 particle connected with particle $({\bf j},\mu)$ via a red (blue) bond [see Fig. \ref{models}(b)].
For each logical qubit, there is a cube associated with a conserved quantity
\begin{eqnarray}
W_{\bf j} &=& \sigma^x_{{\bf j},1} \sigma^x_{{\bf j},2} \sigma^x_{{\bf j},3} \sigma^x_{{\bf j},4} \sigma^z_{{\bf j'}+{\bf e_2},\rm{nb}({\bf j},1)}
\notag \\ &\times &
\sigma^z_{{\bf j'}+{\bf e_1},\rm{nb}({\bf j},2)} \sigma^z_{{\bf j'}-{\bf e_2},\rm{nb}({\bf j},3)} \sigma^z_{{\bf j'}-{\bf e_1},\rm{nb}({\bf j},4)},
\end{eqnarray}
where ${\bf e_1}$ and ${\bf e_2}$ correspond to two unit vectors in a 3D Cartesian coordinate system.
In addition, different $W_{\bf j}$'s commute with each other and also with the Hamiltonian, i.e., $[H^{\rm{2D}}, W_{\bf j}]=0$.

This model is nonintegrable. Thus, we cannot  obtain the exact analytical energy gap.
However, using standard perturbation technique \cite{Stephen_2006}, we arrive at an effective Hamiltonian
$H_{\rm{eff}} ^{\rm 2D}={\rm const.} - (\lambda^6/1536 J^5) \sum_{\bf j} W_{\bf j},$
with an approximate energy gap of
$\Delta E^{\rm 2D} \simeq \lambda^6 /768 J^5$.
We anticipate a larger gap for larger $\lambda$, even if the perturbation is no longer valid.
The 2D cluster state is then obtained following the same preparation procedure as the 1D cluster state.

\begin{figure}[t]
\centering
\includegraphics[scale=0.4]{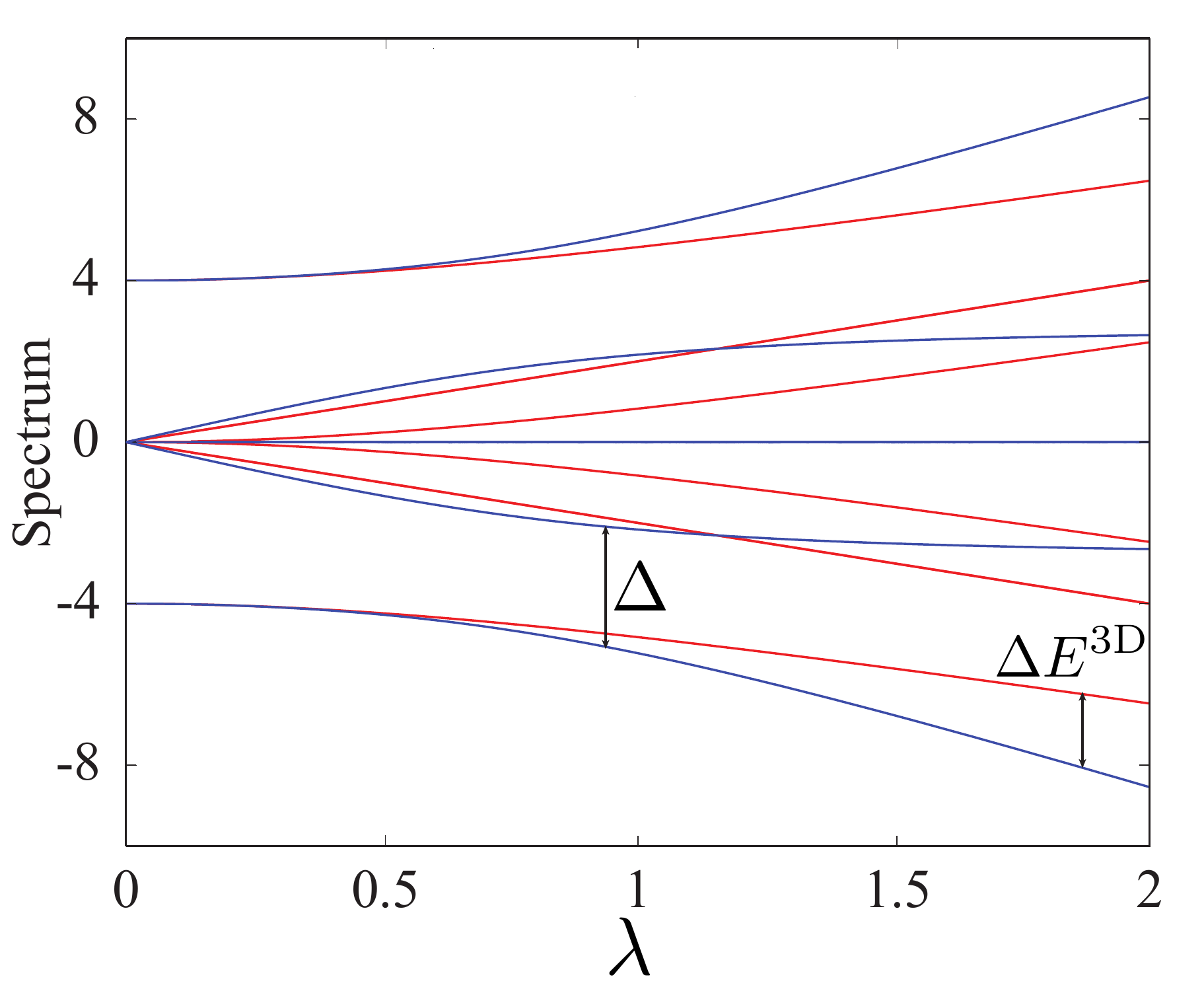}
\caption{Energy spectrum of the 3D square lattice Hamiltonian $H_j ^{\rm 3D}$ versus the coupling $\lambda$.
Energy eigenstates with eigenvalue +1 (eigenvalue -1) of the stabilizer, Eq. (\ref{3D_stabilizer}), are plotted in blue (red) solid lines.
}
\label{spectrum}
\end{figure}

\textit{3D square lattice model}.---A resource state in three dimensions is highly desirable since there exists fault-tolerance quantum error correction (FTQC) scheme \cite{FTQC1,FTQC2} that can be used to correct for any inadvertent error during the computations.
The 3D square lattice model [see Fig. \ref{models}(c)] has the Hamiltonian, $H^{\rm{3D}}= H_0^{\rm 3D} + \lambda V^{\rm 3D}$ where
$
H_0^{\rm 3D}  = -J \sum_{j} \sum_{\langle\mu, \mu ' \rangle} \sigma^z _{(j, \mu)} \sigma^z _{(j, \mu')},
$
and
$
V^{\rm 3D} = - \sum_{j} \sum_{\mu =1}^4 \sigma_{(j, \mu)}^x  \sigma_{\rm{nb}(j, \mu)}^z .
$
Each logical qubit $j$ is encoded in four spin-1/2 particles and, there exists an exact ground state after controlled-phase transformation ($\mathcal{CZ}$) on every bond \cite{Griffin_2008, Klagges_2012}, i.e.,
$
\mathcal{H}^{\rm{3D}}= (\mathcal{CZ})H^{\rm{3D}}(\mathcal{CZ})=\sum_{j} H^{\rm{3D}}_j,
$
where $H^{\rm{3D}}_j= [-J \sum_{\langle\mu, \mu' \rangle} \sigma^z _{(j, \mu)} \sigma^z _{(j, \mu')} -\lambda \sum_{\mu=1}^4 \sigma_{(j,\mu)}^x ]$.
For each plaquette $j$, we notice there exists local conserved quantities $W_j$:
\begin{equation}
W_j ^{\rm{loc}}=(\mathcal{CZ})W_j (\mathcal{CZ})=\prod_{\mu=1}^4 \sigma_{(j,\mu)}^x ,\label{3D_stabilizer}
\end{equation}
such that $W_j ^{\rm{loc}}$'s commute with each other as well as with the Hamiltonian, i.e., $[H^{\rm{3D}} _j,W_j ^{\rm{loc}}]=0$.
Since each $j$ plaquette is independent of each other, we have
$\Delta E^{\rm 3D}= 2\sqrt{ 2J^2 +2\lambda^2 +2\sqrt{J^4+\lambda^4} } -2\sqrt{ J^2 +\lambda^2 } -2J$, the energy gap \cite{Klagges_2012} between its unique ground state and first excited state.
This energy gap, [see Fig. \ref{spectrum}], ensures cooling the system to its unique ground state.
This initial ground state of the time-dependent Hamiltonian $H^{\rm{3D}}_j (\lambda(t))$ remains an approximate ground state of the Hamiltonian throughout the entire evolution as long as the rate of change of $\lambda$ is sufficiently slow satisfying the adiabatic condition \cite{Farhi_2000}.
We also note that the stabilizers $W_j ^{\rm loc}$, Eq. (\ref{3D_stabilizer}), stabilize the instantaneous ground state throughout the adiabatic evolution ($t:0\rightarrow \tau$) since $[H_j ^{\rm 3D},W_j ^{\rm loc}]=0$.
Moreover, there exists a larger energy gap $\Delta$ within the subspace with eigenvalue $+1$ of the stabilizer that in turn allows us to apply a constant $\lambda$ switching rate even though $\Delta E^{\rm 3D}\rightarrow 0$, where $H^{\rm 3D}$ has many degenerate ground states. 
Thanks to the adiabatic evolution and the local stabilizers $W_j ^{\rm{loc}}$, we can concentrate our initially prepared ground states to computationally useful cluster states.

\begin{figure}[t]
\centering
\includegraphics[scale=0.5]{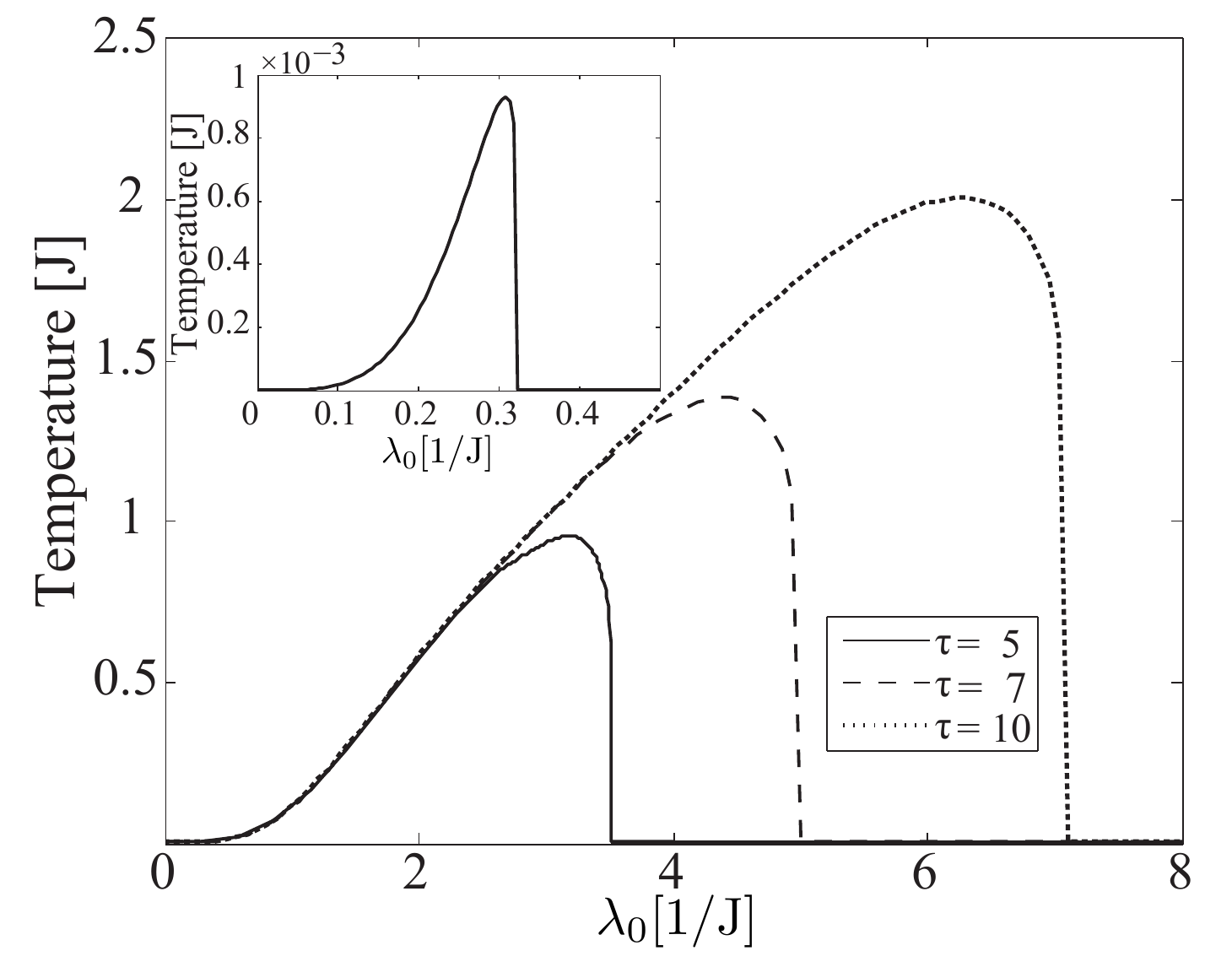}
\caption{
Phase diagram which shows, in temperature and initial coupling $\lambda_0$ space, the lines below which an initial thermal state after the adiabatic evolution is a resource state for fault-tolerent MBQC because the areas enclosed by the lines represent regions with less than 3\% total phase-flip errors while the outer areas are regions with more than 3\% total phase-flip errors.
Solid line in the inset figure corresponds to the ground state without evolution while solid, dashed and dotted lines correspond to the ground state with evolution time $\tau=$ 5, 7, and 10 respectively.
}
\label{Zerror}
\end{figure}

\textit{Error and feasibility of our proposal}.---In the numerical simulation of the 3D cluster state concentration process, we consider a Hamiltonian of the form
$
H^{\rm 3D}_j (t)= H_{j 0} ^{\rm 3D} +  \lambda(t) V^{\rm 3D}_j,
$
where $\lambda \rightarrow 0$ as $t:0\rightarrow \tau$ ($\tau=\lambda_0 /v$ and $\lambda(t) =\lambda_0 -vt$). 
We prepare a thermal state as the initial state,
$
\rho(0)=\mathcal{Z}^{-1}e^{-H^{\rm 3D}_j (0)/T},
$
where $\mathcal{Z}=\text{tr} e^{-H^{\rm 3D}_j (0)/T}$ and $T$ is the temperature of the system setting the Boltzmann's constant to unity.
The ground state without evolution [see Fig. \ref{Zerror} (inset)] corresponds to the case of a perturbational technique \cite{Stephen_2006}.
With our proposal, the three evolution times $\tau=$ 5, 7, and 10 give rise to 3 orders of magnitude higher operating temperature compared to the no-evolution case.
An important observation adduced from Fig. \ref{Zerror} is that the longer the evolution time, the higher the temperature, at which the system ground state can be prepared and the larger phase space region where standard fault-tolerant error correction schemes can be implemented  to correct for possible errors \cite{Supplemental}.

\textit{Discussions}.---Our proposal is not limited to the three models discussed so far and it can also be applied to other models such as the Bartlett and Rudolph's 2D hexagonal lattice \cite{Stephen_2006}, and the Kitaev's 2D honeycomb model \cite{Kitaev_2006}. However, the error correction threshold for the 2D cluster states \cite{Raussendorf_2009,1DFTQC} is believed to be much lower than that of 3D cluster states \cite{FTQC1,FTQC2}. Thus, we are more interested in the implementation of our protocol in generating 3D cluster states. Our proposal benefits from an energy gap protection similar to that of the AKLT resource state \cite{Brennen2008} since the interactions can be switched off sequentially \cite{Supplemental}. Also, our models have a close connection with condensed matter models. 

In summary, we have proposed a means to create cluster states of spin-1/2 particles with just nearest-neighbor two-body interactions via adiabatic evolution, which could be realized with existing technology \cite{Kim2010}.

\begin{acknowledgments}
We acknowledge support from the National Research Foundation \& Ministry of Education, Singapore.
\end{acknowledgments}

\onecolumngrid
\appendix
\section{Error and feasibility of our proposal}
\begin{figure}[htbp!]
\centering
\includegraphics[scale=1]{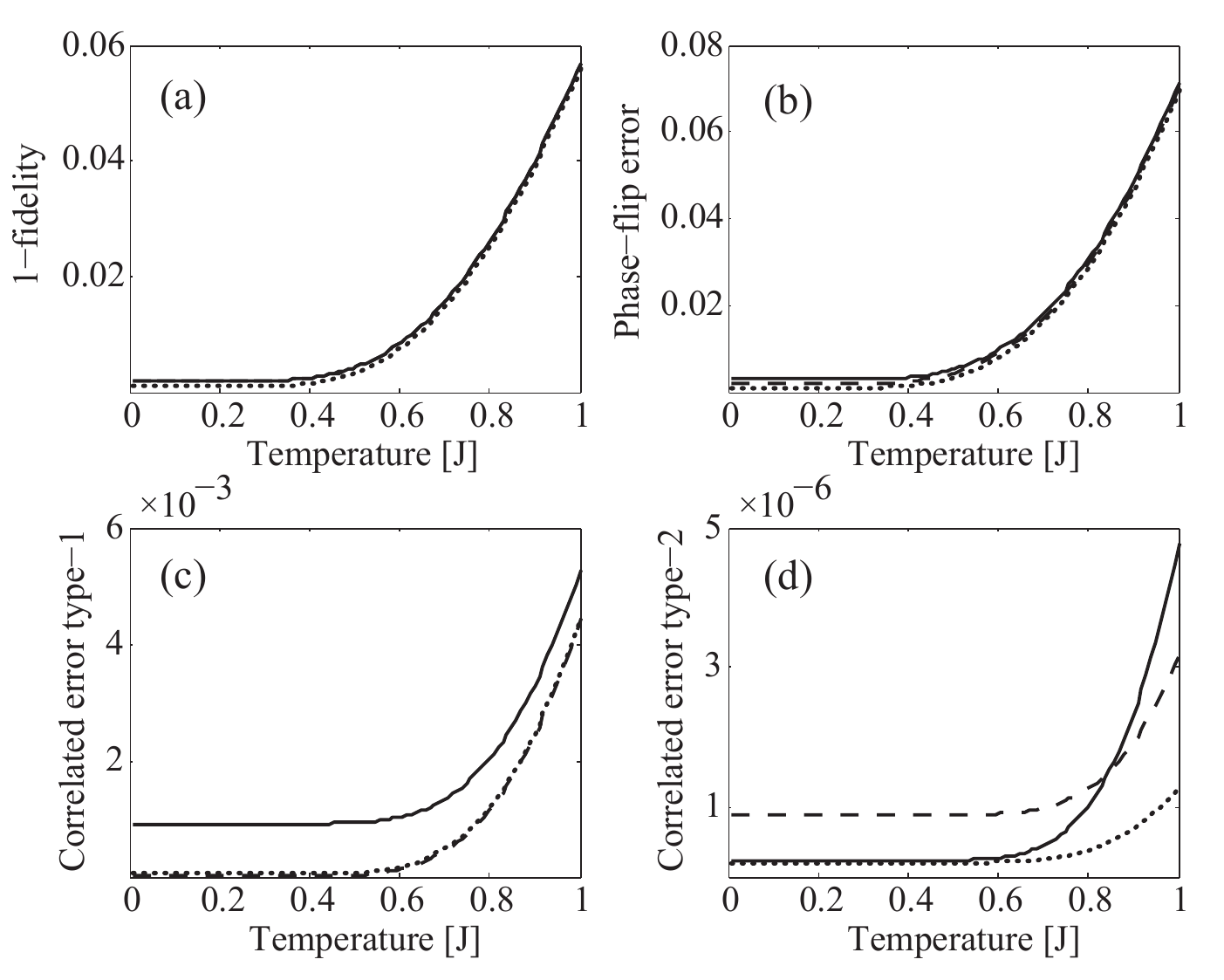}
\caption{(a) Imperfection (1-fidelity) versus temperature plot. 
(b) Total phase-flip error ($E_{\zeta}$) versus temperature plot. 
(c) Correlated error type-1 ($P_{C_1}$) versus temperature plot. 
(d) Correlated error type-2 ($P_{C_2}$) versus temperature plot. 
Solid, dashed and dotted lines represent the evolution with $\tau=$ 5, 7 and 10 respectively with $\lambda_0 =2.5$.  
}
\label{phases_extra}
\end{figure}
\begin{figure}[htbp!]
\centering
\includegraphics[scale=0.4]{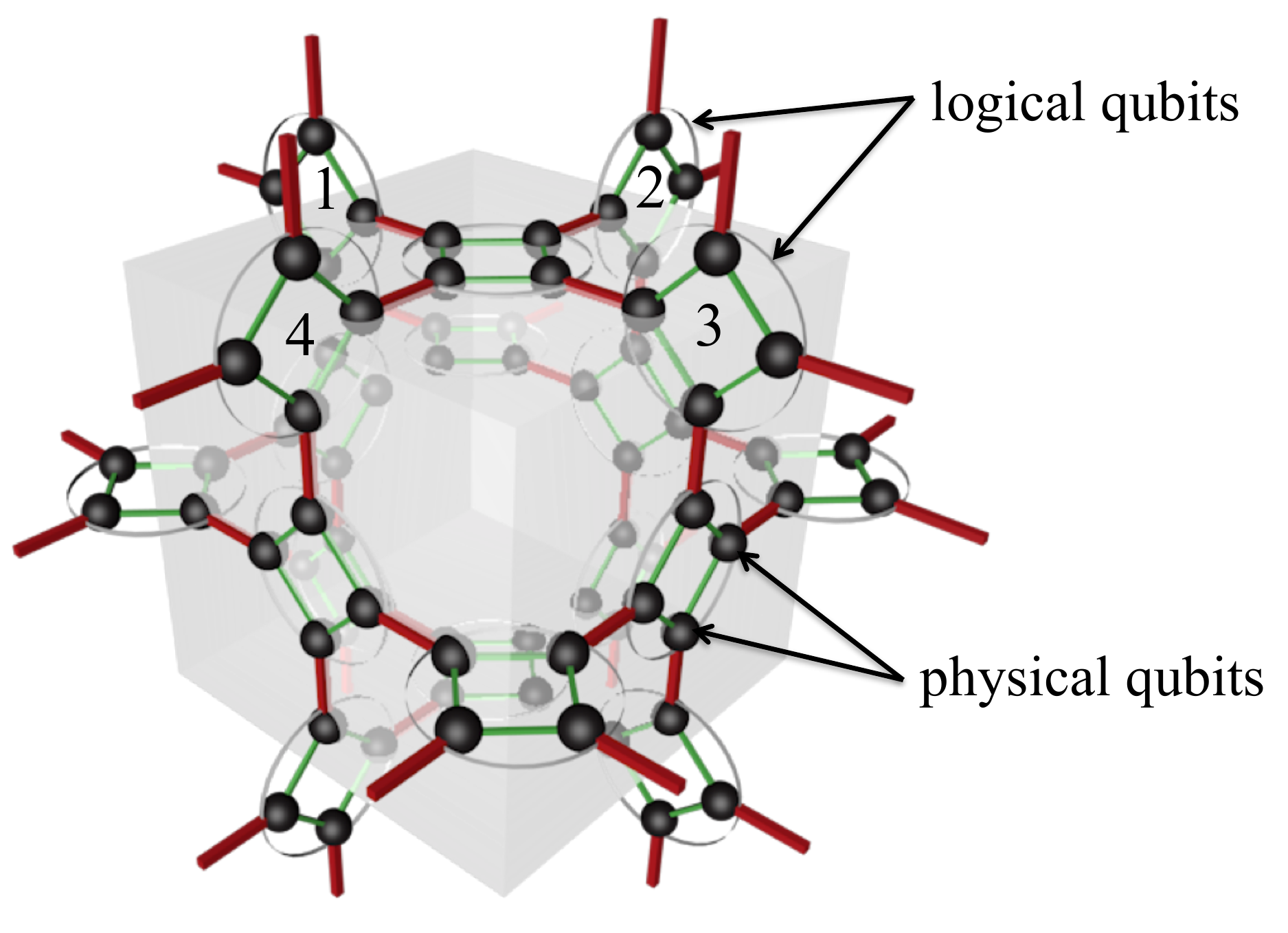}
\caption{An elementary cubic lattice in the three-dimensional square lattice model.
A grey circle of four spin-1/2 particles connected by four green bonds denotes a logical qubit.
Numbers 1, 2, 3 and 4 label four logical qubits located on the four different edges surrounding the central logical qubit sitting at the top face of the cube.
Detailed explanation of the model can be seen in the main text.
}
\label{3D_model_transparent}
\end{figure}

It is well-known that the topological fault-tolerant quantum computing can be used to apply quantum error correction during the course of quantum computations as long as the average total phase-flip errors $E_{\zeta}=P_Z + 4P_{C_1}+2P_{C_2}$ of individual logical qubit in the 3-D model is 3\% or less \cite{FTQC1,FTQC2}. 
We study the adiabatic evolution for various $\tau$'s numerically and estimate the optimal operating temperature of the 3-D model for various combinations of $\tau$ and $\lambda_0$ needed to yield the final states with less than 3\% phase-flip errors $E_{\zeta}$ relevant for the designated cluster states.
The results are shown in Fig. 3 of the main text.
Errors on each square lattice in the final state can be expressed by a superoperator
\begin{equation}
E = F[\mathbb{I}] + P_Z \left(\sum_{k=1}^4 [Z_k]\right) +
\frac{P_{C_1}}{2} \left(\sum_{m\leftrightarrow l=1}^4[Z_l Z_m]\right) +\frac{P_{C_2}}{2} ([Z_1 Z_3] + [Z_2 Z_4]),
\label{3D_errors}
\end{equation}
where $F$ refers to fidelity, $P_Z$ refers to local phase-flip errors, $P_{C_1}$ refers to correlated errors type-1, $m \leftrightarrow l$ means the sites $m$ and $l$ are graphically connected, $P_{C_2}$ refers to correlated errors type-2 [see Fig. \ref{3D_model_transparent}] and a superoperator satisfies $\mathcal{O}[\rho]=\mathcal{O}\rho\mathcal{O}^\dagger$.

The efficiency and effectiveness of the fault tolerance quantum computation depend not only on $P_Z$ (estimated in the main text) but also on correlated errors $P_{C_1}$ and $P_{C_2}$ among neighboring logical qubits.
Here, we demand $P_{C_1} + P_{C_2}\ll P_Z$ so that the conclusion that we have adduced from Fig. 3 in the main text is valid.
From the numerical evidence shown in Fig. \ref{phases_extra} (b-d), it is clear that the above mentioned requirement is fulfilled in all the three cases.
Moreover, we see that halving temperature from $T=1\rightarrow 0.5$ for $\tau=5,7,10$ reduces the total phase-flip errors by about one order of magnitude [see Fig. \ref{phases_extra}(b)].

\section{Sequential adiabatic switch-off}
\begin{figure}[htbp!]
\centering
\includegraphics[scale=0.63]{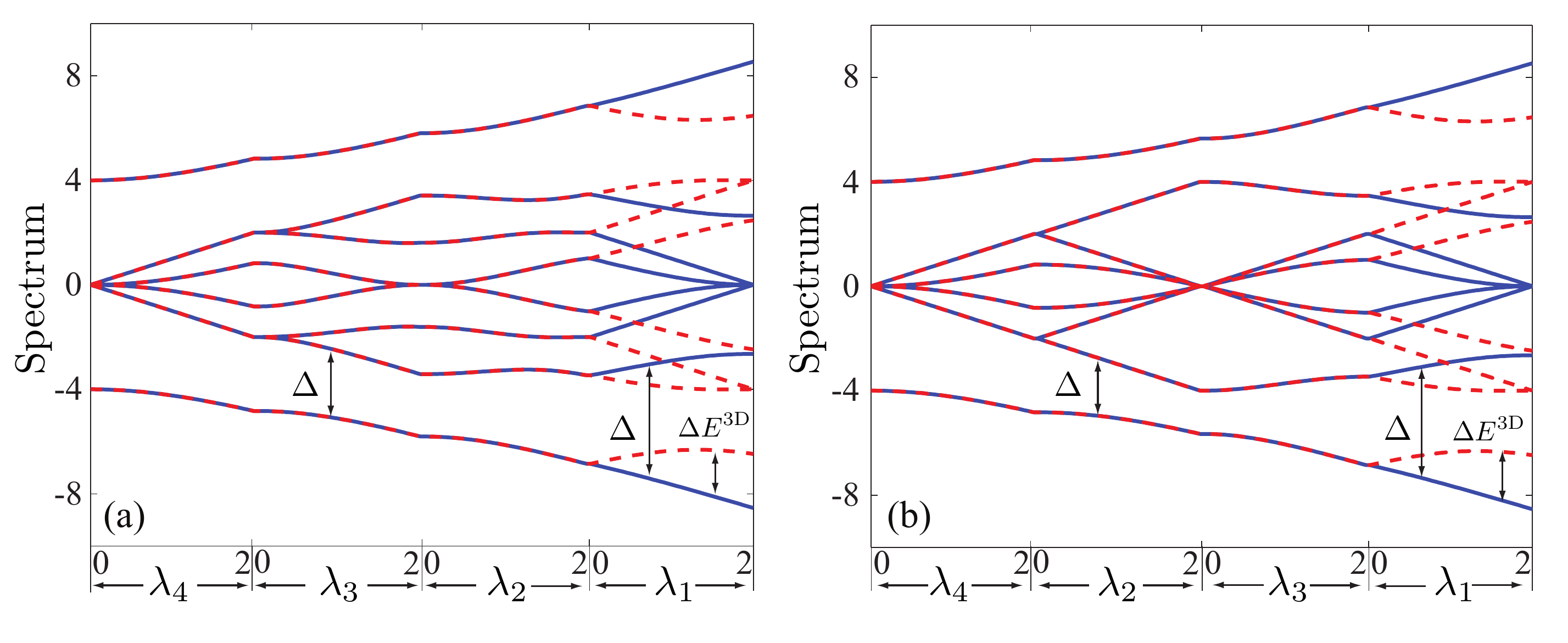}
\caption{
Energy spectrum of the 3-D square lattice Hamiltonian $H_j ^{\rm 3D}$ versus the coupling $\lambda$'s while (a) the surrounding logical qubit-1, 2, 3, and 4 and (b) the surrounding logical qubit-1, 3, 2, and 4 [see Fig. \ref{3D_model_transparent}] are disconnected in sequential adiabatic manner where $\lambda_1$, $\lambda_2$, $\lambda_3$ and $\lambda_4$ are coupling constants between the central $j$th logical qubit and the surrounding logical qubit-1, 2 ,3 and 4 respectively.
Energy eigenstates with eigenvalue +1 (eigenvalue -1) of the stabilizer are plotted in blue solid (red dashed) lines.
Each $\lambda$ is adiabatically tuned from 2 to 0 as in Fig. 2 of the main text.
}
\label{spectrum_onebyone}
\end{figure}

The 3-D system is initially prepared or cooled down to its unique ground state, which is not a computational resource state.
We then adiabatically switch off $\lambda_{(j,\mu)}$'s for each logical qubit $j$ one at a time.
By doing so, we drive the $j$th logical qubit state into a computational resource state at the end of the adiabatic evolution.
We then perform measurement onto this resource state.
After the measurement of the $j$th logical qubit, Hamiltonian of the fully connected logical qubits in the residual Hamiltonian $H^{\rm 3D}_{\rm {res}}= \sum_k ^{N-5} [-J\sum_{\mu \leftrightarrow \mu'} \sigma_{(k,\mu)}^z \sigma_{(k,\mu')}^z - \sum_{\mu=1}^4 \lambda_{(k,\mu)} \sigma_{(k,\mu)}^x \sigma_{nb(k,\mu)}^z ]\rm{(fully\hspace{.1cm}connected)}$+$\sum_{m\in nb(j)}^4[-J \sum_{\xi \leftrightarrow \xi'} \sigma_{(m,\xi)}^z \sigma_{(m,\xi')}^z -\sum_{\xi, (m,\xi)\nleftrightarrow (j,\mu)}^3 \lambda_{(m,\xi)} \sigma_{(m,\xi)}^x \sigma_{nb(m,\xi)}^z ] \rm{(partially\hspace{.1cm}connected)}$, is still gapped (with $\Delta E ^{\rm 3D}$) as before and the four neighboring partially connected ones are no more protected by the gap.
Here, $N$ is the total number of logical qubits, $(m,\xi)\nleftrightarrow (j,\mu)$ means the physical qubit $\xi$ belonged to the $m$th logical qubit is not graphically connected to the physical qubit $\mu$ belonged to the $j$th logical qubit and $nb(j)$ means neighbour of the $j$th logical qubit.
From this observation, we draw attention that after every consumption of a resource state, there could be some other partially connected logical qubits located on a boundary between measured logical qubits and fully connected ones.
These partially connected ones should be measured immediately or treated as redundant and discarded.
Moreover, adiabatically switching off $\lambda$'s of the $j$th qubit from the bulk does not couple instantaneous ground states with excited states because there is no level crossing in the energy spectra (see Fig. \ref{spectrum_onebyone}), and this can be done monotonically in time due to the presence of larger energy gap $\Delta$ in the subspace defined by stabilizers throughout the entire adiabatic evolution.
To be specific, from Fig. \ref{spectrum_onebyone}, we note that the ground state subspace with +1 eigenvalue of the stabilizers has non-zero energy gap $\Delta >\Delta E^{\rm 3D}>0$ when $\lambda$'s are being adiabatically turned off in succession.

With all the properties described above, our model does enjoy energy gap protection similar to the AKLT resource state \cite{Brennen2008} while the eminent advantage with our proposal is that we are able to create cluster states of spin-1/2 particles with just nearest-neighbor two-body interactions.

\newpage

\twocolumngrid

\end{document}